# Frequency-swept Brillouin spectroscopy

**LIN KANG, SHAHRIN JAHAN JAIMA, JITAO ZHANG***

*Department of Biomedical Engineering, Institute for Quantitative Health Science & Engineering, Michigan State University, East Lansing, MI 48824, USA*

**zhan2399@msu.edu*

**Spontaneous Brillouin microscopy provides non-contact access to the viscoelastic properties of materials. Standard measurements typically utilize a single-wavelength continuous-wave laser to illuminate a specimen and excites a spontaneous Brillouin scattering signal, which is then recorded by a spectrometer. Existing Brillouin spectrometers are mostly built around either a scanning Fabry-Pérot etalon or a virtually imaged phased array (VIPA) etalon. While performing well, these setups are bulky and demand significant optical expertise to construct and maintain. Here, we propose a new approach to conducting Brillouin spectroscopy. In this approach, a frequency-swept laser excites a series of Brillouin spectra with incrementally shifted central frequencies, while a frequency picker, consisting of a narrow bandpass filter and a highly sensitive single-photon detector, sequentially records each spectral component to reconstruct the full spectrum. We demonstrate that the frequency-swept Brillouin spectroscopy can run under shot-noise limited condition using standard samples. Compared with conventional spectrometers, our system features a much more compact design for portable applications and holds potential for rapid mechanical imaging through multiplexing.**

Spontaneous Brillouin light scattering provides a label-free, non-contact access to the viscoelastic properties of soft and biological materials [1–3]. Because the measured Brillouin frequency shift and linewidth directly link to the complex longitudinal modulus of the probed volume, Brillouin microscopy has become a valuable tool for probing the biomechanical properties of live cells [2,4], developing tissues [5,6], and engineered biomaterials [7]. Standard measurements of Brillouin spectrum typically utilize a single-wavelength continuous-wave laser with narrow linewidth to illuminate a specimen and excite a Brillouin scattering signal, which is then recorded and analyzed by a dedicated spectrometer (Figure 1a). The great majority of existing Brillouin spectrometers are built around a spatially dispersive device [8,9], typically a virtually imaged phased array (VIPA) etalon [10] or a scanning Fabry-Pérot (FP) etalon [11,12]. Currently, the multi-stage VIPA-based spectrometer, and the tandem multi-pass FP etalon-based spectrometer represent the state-of-the-art for Brillouin spectroscopy. While these spectrometers perform well, their sophisticated optical configurations require precise alignment of numerous components, resulting in bulky setups that demand substantial optical expertise for construction and routine maintenance.

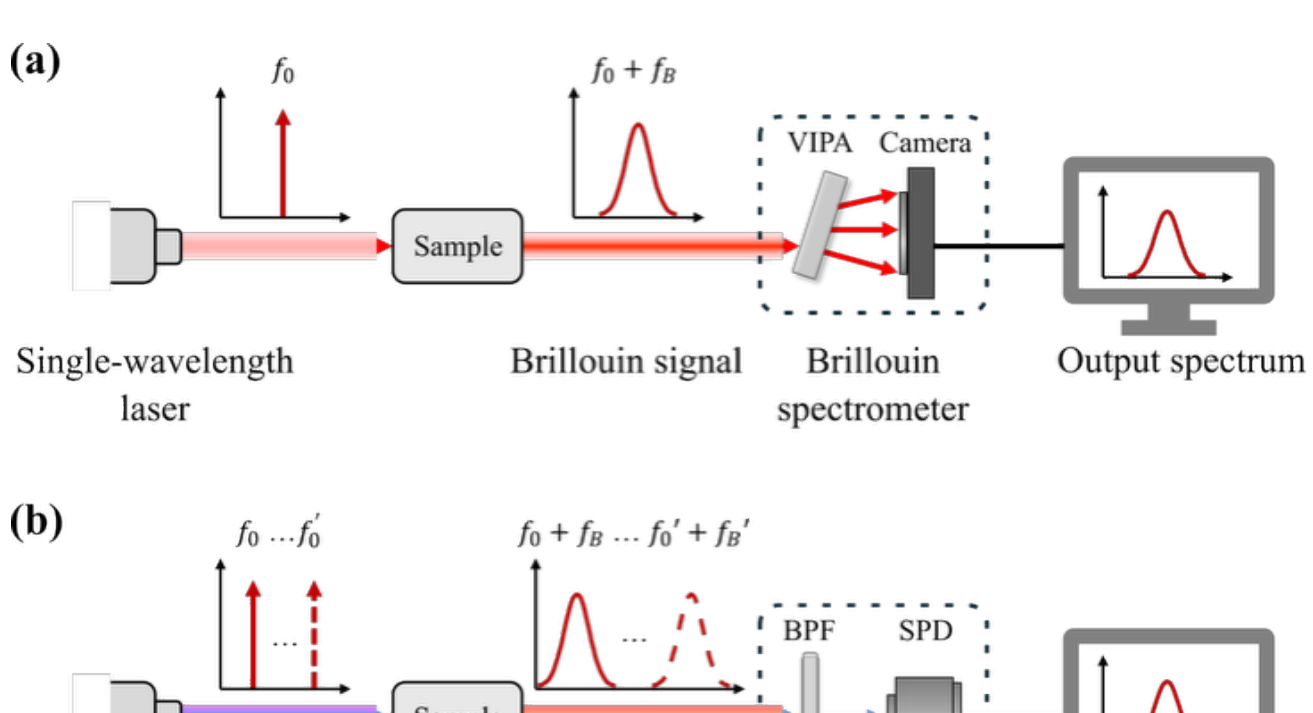

Fig. 1. Concept of frequency-swept Brillouin spectroscopy. (a) Conventional VIPA-based Brillouin spectroscopy: a single-frequency laser ($f_0$) excites the sample, and the Brillouin-shifted signal ($f_0 + f_B$) is angularly dispersed by a VIPA onto a camera. (b) Frequency-swept Brillouin spectroscopy: the excitation frequency is swept ($f_0 \rightarrow f_0'$) while the correspondingly shifted signal is passed through a narrow bandpass filter (BPF) and recorded by a single photon detector (SPD), reconstructing the spectrum in time domain.

Here, we propose a different strategy for Brillouin spectroscopy by encoding the spectrum in time instead of dispersing it spatially (Figure 1b). In our approach, a frequency-swept laser is used to excite a series of Brillouin spectra with incrementally shifted central frequencies, while a frequency picker, including an ultra-narrow bandpass filter and a highly sensitive single-photon detector, is used to sequentially record each spectral component to reconstruct the full spectrum. This frequency-swept Brillouin spectrometer (FSBS) is very compact in size and easy for construction and maintenance. Our idea is inspired by the recent advances in swept-source Raman spectroscopy [13,14], which has allowed high-efficiency detection of molecular fingerprint. Unlike Raman scattering, the Brillouin frequency shift relies on the excitation wavelength

$$\omega_B = \frac{2n}{\lambda} V \sin\frac{\theta}{2}, \qquad (1)$$

where $n$ is the refractive index of the sample, $\lambda$ is the incident laser wavelength, $V$ is the sound velocity, and $\theta$ is the scattering angle.

However, sweeping the frequency of the laser source by ~10 GHz will only affect the Brillouin shift by less than $3\times10^{-5}$, which is well below the relative spectral precision of the instrument (e.g., ~$10^{-3}$) and thus can be neglected. In addition, because the shift and linewidth of a Brillouin spectrum is orders of magnitude smaller than that of a Raman spectrum, reconstructing the spectrum demands filters with ultra-narrow bandwidth (e.g., on the order of ~100 MHz or less), which is beyond the capability of commercial filters designed for Raman spectroscopy and requires custom design.

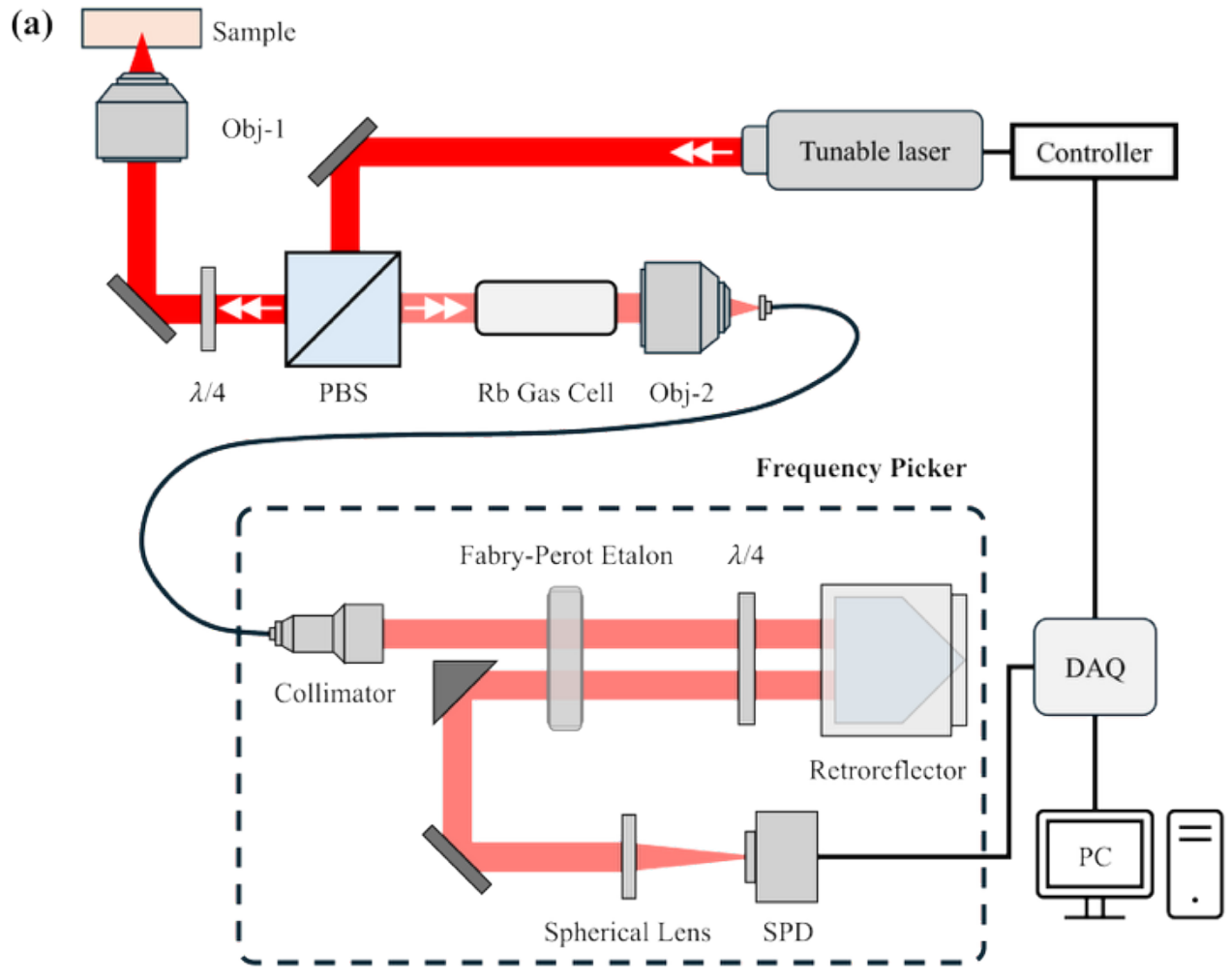

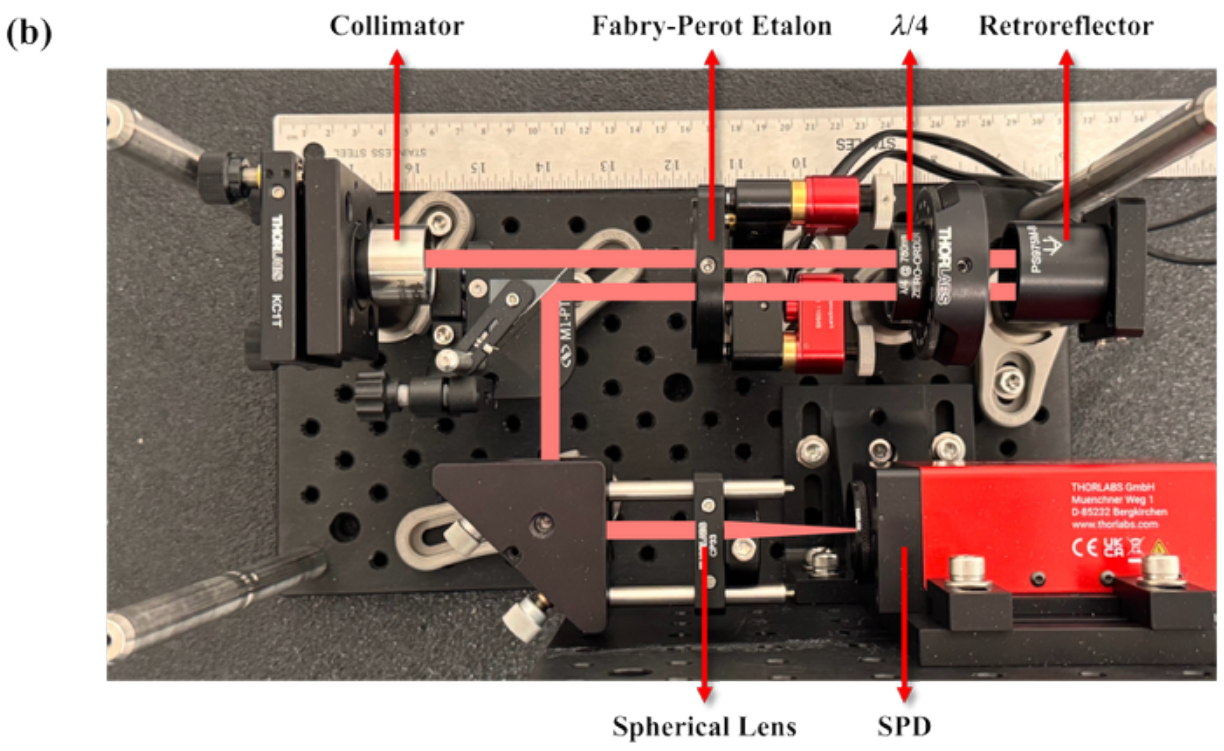

Fig. 2. (a) Optical layout of the frequency-swept Brillouin spectroscopy. Obj-1, Obj-2: objective lenses; $\lambda/4$: quarter-wave plate; PBS: polarizing beam splitter; Rb: rubidium vapor cell. DAQ: Data Acquisition Board. Dashed box: frequency-picker module, comprising a fiber collimator, double-pass Fabry-Pérot etalon with retroreflector, focusing lens, and single photon detector (SPD). (b) Photograph of frequency picker from the dashed box in (a), with red lines highlighting the beam pathway.

The schematic of the optical layout is shown in Figure 2a. The light source is a 780-nm tunable laser (DL pro, Toptica) controlled through a computer interface. Its spectrum is cleaned using a pair of Bragg filters (BP-780, OptiGrate). To establish an absolute frequency reference, the laser is first locked to the hyperfine absorption lines of a rubidium ($^{85}$Rb) vapor cell. During measurement, the laser is unlocked and its frequency continuously swept by driving the piezoelectric actuator with a ramp voltage, synchronized with the data acquisition through a voltage-to-frequency scale factor of 0.3839 GHz/V. The linearly polarized laser light passes through a polarizing beam splitter (PBS) and a quarter-wave plate ($\lambda/4$) to generate circular polarization before being focused into the sample by a 20×/0.45 NA objective lens (Obj-1). Backscattered Brillouin signal, collected by the same objective lens, passes back through the quarter-wave plate and emerges with orthogonal linear polarization, causing it to be transmitted by the PBS toward the detection path. A hot Rb vapor cell with a length of 150 mm is used to suppress excessive Rayleigh scattering. The signal is then coupled into a multimodal fiber using a second objective lens (Obj-2) and delivered to the frequency picker module (dashed box; photo of the setup is provided in fig. 2b) for spectral analysis. Inside the frequency picker, the beam from a fiber collimator is transmitted through a double-pass FP etalon (FSR = 15 GHz, LightMachinery), serving as a narrow bandpass filter, and then focused onto a single-photon detector (SPD) (SPDMH2, Thorlabs) by a spherical lens. To maximize signal transmission, a quarter-wave plate is used to compensate for the minor polarization change produced by total internal reflection within the retroreflector. The transmission band of the filter is continuously tunable by tilting the solid etalon using a precise motorized mount (8816-8, Newport). A Python-based software interface is developed to synchronizes the laser-frequency sweep with photon counting by the SPD. Specifically, a high-speed data acquisition (DAQ) board time-tags photon arrival events relative to the laser sweep voltage to build a time-resolved photon-count histogram for each sweep. As shown in Fig. 2b, the frequency picker module has a dimension of about 30 × 15 × 15 cm, which is several times smaller than a standard VIPA-based spectrometer.

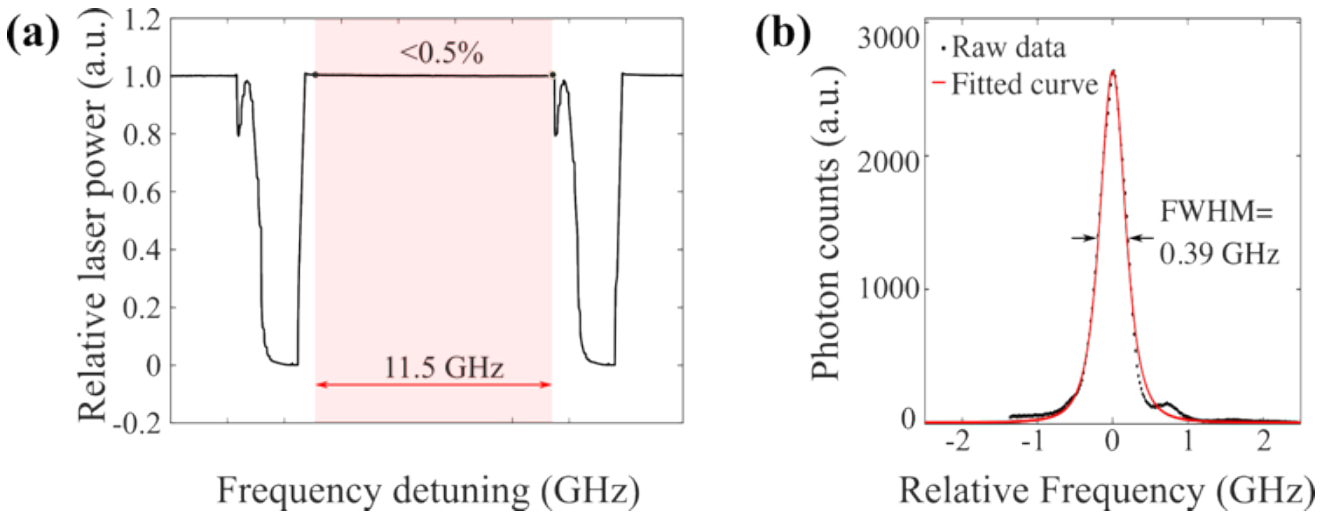

Fig. 3. Instrument characterization on the unshifted laser line. (a) Relative excitation power versus frequency detuning; power is stable to <0.5% across an 11.5 GHz window (shaded). (b) Spectral response of the frequency picker module: raw photon counts (black) and Airy$^2$ fit (red), FWHM = 0.39 GHz.

Figure 3a shows the power stability of the laser source during frequency sweeping. Over an 11.5 GHz frequency sweeping range, the power fluctuation is less than 0.5%, suggesting negligible power variation across the sweep. To characterize the bandwidth of the frequency picker, we used the laser light as a monochromatic input (linewidth < 0.3 MHz), giving the resulting spectral response shown in Fig. 3b. The spectrum is fitted by an Airy$^2$ function, yielding a linewidth (Full-width-at-Half Maximum, FWHM) of 0.39 GHz with a total optical throughput of approximately 32%.

For sample measurement, we set the frequency sweeping range to 0.69 GHz, which is constrained by the absorption bandwidth of the Rb vapor cell used for assisting Rayleigh suppression, rather than by any intrinsic limitation of the FSBS. For example, replacing the double-pass FP etalon with a high-extinction tandem FP etalon for filtering would eliminate the need of the Rb vapor cell [12]. However, this limited range prevents direct measurement of the absolute Brillouin shift from frequency sweeping alone. To establish a frequency reference, the laser is first locked to the Rb gas cell while the FP etalon is physically tilted to align its transmission peak with the expected Brillouin signal frequency. The laser is then unlocked and swept across the 0.69-GHz window to acquire the Brillouin spectrum. Because this sweeping window captures only the truncated Brillouin spectrum, we adopt a two-step baseline-anchored fitting method to extract the peak's position. First, the Brillouin spectrum is recorded with the FP etalon at its optimal transmission angle. The etalon is then slightly deflected to its minimum transmission position to acquire an independent background trace under identical conditions. Each sample's spectrum is then fitted to a Lorentzian profile anchored at the measured background floor, from which the Brillouin peak position and its localization precision are determined across repeated sweeps. We validate this fitting method by simulating truncated FSBS spectra of water using the full spectra data acquired by the VIPA-based spectrometer. Gradually restricting the width of the fitting window on the VIPA data confirmed that the extracted Brillouin shift remains stable as long as the peak apex is captured within the sweeping window (Supplement 1, Section S1).

Next, we evaluate the performance of the FSBS system using standard reference materials, methanol and water, whose backward scattering Brillouin shifts at 780 nm under room temperature are well documented (i.e., methanol: 3.80 GHz, water: 5.09 GHz). The input laser power is 13.5 mW, and the frequency sweeping rate changes from 0.1 Hz to 40 Hz, with 10 spectral points recorded per sweep. Representative Brillouin spectra of methanol and water acquired by the FSBS are shown in Figs. 4a and 4b. For methanol, the sweep window is larger than its FWHM linewidth, allowing reliable determination of both peak location and linewidth. For water, the sweeping window is smaller than the FWHM linewidth. While the localization of the peak remains accurate, the linewidth cannot be faithfully extracted and thus should be discarded. The precision of the peak localization was quantified over 50 consecutive sweeps per sample at a sweeping rate of 0.5 Hz. The calculated standard deviations are 5.0 MHz for methanol and 9.6 MHz for water, respectively.

For cross-validation, the samples were measured side-by-side using a single-stage VIPA-based spectrometer under identical conditions, with only the detection fiber switched from the FSBS setup to the VIPA-based spectrometer. A detailed schematic of the VIPA-based spectrometer is provided in the Supplement 1 (Section S2). Briefly, the collected light from a fiber collimator is coupled into the entrance window of a VIPA etalon (FSR=15 GHz, LightMachinery), which spatially disperses optical frequencies into angularly separated beams. A pair of cylindrical lenses then project the spectral pattern onto a slit, which is used to physically block any residual Rayleigh light. The pattern is then reimaged on to an EMCCD camera (iXon Life897, Andor) by a lens pair. For each sample, 100 consecutive frames were captured at 50 ms exposure time per individual frame. Each frame was fitted with a Lorentzian function to extract the Brillouin shift and linewidth, and the reported values representing the mean value across all 100 frames. The resulting spectra are shown in Fig. 4c and 4d, with measured Brillouin linewidths of 0.466 GHz for methanol and 0.592 GHz for water.

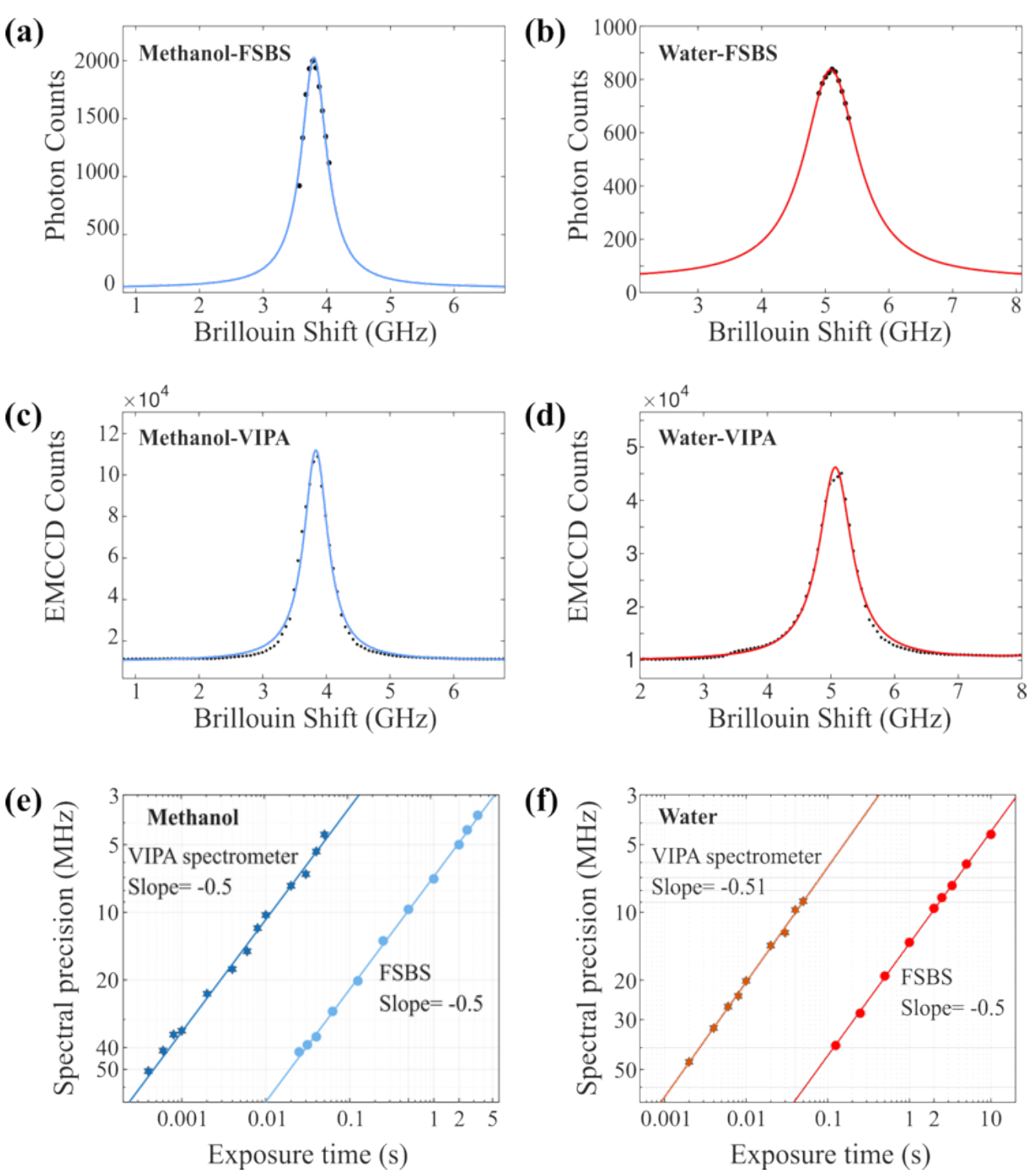

Fig. 4. Performance characterization of the frequency-swept Brillouin spectroscopy. Reconstructed Brillouin spectra of (a) methanol and (b) water measured by the FSBS, anchored to their accepted Brillouin shifts of 3.80 GHz and 5.09 GHz, respectively. Brillouin spectra of (c) methanol and (d) water acquired by a conventional VIPA-based spectrometer under identical conditions. Spectral precision versus exposure time for (e) methanol and (f) water, benchmarked against the VIPA-based spectrometer under log-log scale.

To confirm the FSBS system operates at its theoretical limit, we evaluated measurement precision against exposure time of Brillouin spectra at a constant laser power of ~13.5 mW for both samples. Figure 4e and 4f plot spectral precision versus exposure time on a log-log scale. In the FSBS system, the exposure time is defined as the duration of a single frequency sweep (i.e., the reciprocal of the sweeping rate). At each sweeping rate, 50 spectra were acquired, and the standard deviation of the fitted peak locations was used to quantify the spectral precision. Similar datasets were acquired using the VIPA-based spectrometer by adjusting the exposure time of the EMCCD camera. For both samples, the datasets were linearly fitted with a slope of ~0.5, suggesting that both the FSBS system and the VIPA-based spectrometer operate under shot-noise limited condition. This confirms that photon shot noise is the dominant source and that sweep jitter and detector dark counts do not degrade measurement performance under current experimental settings.

At a benchmark precision of 10 MHz, the FSBS system requires an acquisition time approximately 50 times longer than that of the VIPA-based spectrometer (e.g., 500 ms versus 10 ms for methanol, and 2 s versus 40 ms for water). This discrepancy in acquisition speed is expected, given the fact that VIPA-based spectrometer captures the entire Brillouin spectrum in a single shot by spatially dispersing it across the detecting camera, whereas the FSBS system must sample each frequency components sequentially. However, this discrepancy may reverse for widefield or full-field Brillouin imaging. A VIPA-based spectrometer has already used one spatial dimension of a 2D camera for spectrum analysis, leaving only a single free dimension for spatial multiplexing, which is the basis of existing line-scanning Brillouin microscopy [15–17]. The FSBS system, in contrast, encodes the spectrum in time domain rather than in space, leaving both spatial dimensions available for multiplexing. This makes it possible to illuminate and image the full field of view at each swept frequency with a 2D detector array, an idea that has been recently demonstrated in different implementations [18]. Because acquisition time in this configuration is set by the number of frequency steps rather than the number of spatial pixels, the total throughput for imaging a large field of view can easily favor the FSBS system over a confocal or line-scanning VIPA-based spectrometer as the pixel number grows. In addition, the FSBS can be readily implemented on existing tandem multi-pass FP etalon spectrometer [19]. In this case, mechanical scanning of the etalon is replaced by laser frequency sweeping, leaving the spectrometer without any physically moving part and thereby rendering it more stable and compact.

Due to its limited frequency sweep range, the current setup is mostly suitable for relative measurement of Brillouin shifts against a known material (e.g., water). To determine the absolute Brillouin shift, the tilt angle of the FP etalon can be pre-calibrated using reference materials, and a quick coarse scan prior to each experiment will resolve any frequency ambiguity. However, as mentioned earlier, this limitation comes from the Rb vapor cell, which might not be necessary in other implementations.

Detection sensitivity of the FSBS can be further improved by replacing the single photon detector with a superconducting nanowire single photon detector (SNSPD). SNSPDs offer substantially higher quantum efficiency than conventional SPD [14], which would reduce the required dwell time per frequency component and narrow the current speed gap with VIPA-based spectrometers.

More broadly, the concept of frequency-swept Brillouin spectroscopy is a general framework that works with different types of narrow bandpass filters. The double-pass FP etalon used in this work is only for proofing the concept, and other approaches for filtering are equally compatible. For example, laser-induced circular dichroism atomic line monochromator provides sub-GHz bandwidth without relying on angular dispersion and is well suited to full-field detection configuration [20]. In addition, existing tandem multi-pass FP interferometers already available in many laboratories could also be directly repurposed as a narrow bandpass filter.

In conclusion, the FSBS represents a distinct approach to Brillouin spectroscopy compared to existing methods. We demonstrated its feasibility using a proof-of-concept setup. Because FSBS does not require a traditional spectrometer, the instrument can be compact and easy to maintain. Furthermore, since the spectral acquisition is performed in the time domain rather than the spatial domain, FSBS holds great potential for full-field Brillouin imaging via spatial multiplexing.

**Funding.** U.S. National Science Foundation (CBET- 2546314).

**Disclosures.** A provisional patent application related to this research has been filed. The authors declare no conflicts of interest.

**Data availability**. Data underlying the results presented in this paper may be obtained from the authors upon reasonable request.

**Supplemental document.** See Supplement 1 for supporting content.